\documentclass[aps, prd, twocolumn, amsmath, floats, floatfix, superscriptaddress, nofootinbib, showpacs]{revtex4-2}

\usepackage{epsfig}
\usepackage{amsmath,amsfonts,amssymb}
\usepackage[caption=false, justification=centerlast]{subfig}
\usepackage{verbatim}
\usepackage{float}
\usepackage{morefloats}
\usepackage[dvipsnames]{xcolor}
\usepackage[normalem]{ulem}
\usepackage{graphicx}
\usepackage{multirow}
\usepackage{makecell}
\usepackage{tabularx}
\usepackage{cancel}
\usepackage{url}
\usepackage{breakurl}
\usepackage[breaklinks]{hyperref}
\usepackage[T1]{fontenc}
\usepackage[capitalise]{cleveref}
\crefname{section}{Sec.}{Secs.}
\crefname{table}{Tab.}{Tabs.}
\crefname{figure}{Fig.}{Figs.}
\crefname{equation}{Eq.}{Eqs.}
\crefname{appendix}{Appendix\ }{Appendix\ }

\providecommand{\openone}{\leavevmode\hbox{\small1\kern-3.8pt\normalsize1}}

\usepackage[Q=yes,pverb-linebreak=no]{examplep}
\usepackage{aas_macros}
\usepackage{scrextend}

\usepackage{siunitx}

\usepackage{nameref}

\begin{document}

\title{\boldmath 
Machine-Learning Love: classifying the equation of state of neutron stars with Transformers
\unboldmath}

\author{Gon\c{c}alo~Gon\c{c}alves}
\email{goncalo.mota.goncalves@gmail.com}
\affiliation{LIP, Department of Physics, University of Coimbra, 3004-516 Coimbra, Portugal}

\author{M\'arcio~Ferreira}
\email{marcio.ferreira@uc.pt}
\affiliation{CFisUC, 
    Department of Physics, University of Coimbra, 3004-516  Coimbra, Portugal}

\author{Jo\~ao~Aveiro}
\email{joao@aveiro.me}
\affiliation{CFisUC, 
    Department of Physics, University of Coimbra, 3004-516  Coimbra, Portugal}

\author{Antonio~Onofre}
\email{antonio.onofre@cern.ch}
\affiliation{Centro de F\'{\i}sica das Universidades do Minho e do Porto (CF-UM-UP), Universidade do Minho, 4710-057 Braga, Portugal}

\author{Felipe~F.~Freitas}
\email{felipefreitas@ua.pt}
\affiliation{Departamento de F\'\i sica da Universidade 
de Aveiro and \\
Centre  for  Research  and  Development  in  Mathematics  and  Applications  (CIDMA)\\
Campus de Santiago, 3810-183 Aveiro, Portugal}

\author{Constan\c{c}a~Provid\^encia}
\email{cp@uc.pt}
\affiliation{CFisUC, 
    Department of Physics, University of Coimbra, 3004-516  Coimbra, Portugal}

\author{Jos\'e~A.~Font}
\email{j.antonio.font@uv.es}
\affiliation{Departamento de Astronom\'ia y Astrof\'isica, Universitat de Val\`encia,
Dr. Moliner 50, 46100, Burjassot (Val\`encia), Spain}
\affiliation{Observatori Astron\`omic, Universitat de Val\`encia,
Catedr\'atico Jos\'e Beltr\'an 2, 46980, Paterna (Val\`encia), Spain}

\begin{abstract}
The use of the Audio Spectrogram Transformer (AST) model for gravitational-wave data analysis is investigated. The AST machine-learning model is a convolution-free classifier that captures long-range global dependencies through a purely attention-based mechanism. In this paper a model is applied to a simulated dataset of inspiral gravitational wave signals from binary neutron star coalescences, built from five distinct, cold equations of state (EOS) of nuclear matter. From the analysis of the mass dependence of the tidal deformability parameter for each EOS class it is shown that the AST  model achieves a promising performance in correctly classifying the EOS purely from the gravitational wave signals, especially when the component masses of the binary system are in the range $[1,1.5]M_{\odot}$. Furthermore, the generalization ability of the model is investigated by using gravitational-wave signals from a new EOS not used during the training of the model, achieving fairly satisfactory results. Overall, the results, obtained using the simplified setup of noise-free waveforms, show that the AST model, once trained, might allow for the  instantaneous inference of the cold nuclear matter EOS directly from the inspiral gravitational-wave signals produced in binary neutron star coalescences.   
\end{abstract}

\maketitle

\section{Introduction}

Gravitational waves (GWs) are a crucial source of information in understanding the properties of astrophysical compact objects like black holes and neutron stars (NS). In particular, the GWs emitted during the coalescence of binary NS (BNS) carry important information on the high-density behavior of cold NS matter. 
The tidal deformability of NS quantifies the quadrupolar deformation of the star
when subject to an external tidal field, such as the one generated by the companion star in BNS systems. The tidal deformability parameter is a decreasing function of the NS mass, spanning many orders of magnitude, that is sensitive to the equation of state (EOS) of NS matter. The stars’ tidal response affects the binding energy of the BNS system and the rate of emission of gravitational waves. The EOS of NS matter has a direct imprint on the GW signals through the tidal deformability of each binary component~\cite{Read:2013,Chatziioannou:2020pqz,Dietrich:2020eud}. 

The landmark GW observation of the BNS merger event GW170817 by the Advanced LIGO and Advanced Virgo detectors~\cite{LIGOScientific:2017vwq}, followed by dozens of observations of its electromagnetic counterpart GRB179817A/AT2017gfo~\cite{grb,kilo}, gave rise to the new field of multi-messenger astronomy with gravitational waves. 
The observation of the GW170817 event allowed for the inference of the effective tidal deformability~\cite{Flanagan:2008} of the binary system  $\tilde{\Lambda}\le 800$ (90\% confidence, using a low-spin prior) \cite{LIGOScientific:2017vwq}. A follow up reanalysis tightened the constraints to  $\tilde{\Lambda}=300^{+420}_{-230}$ (90\% confidence), under minimal assumptions about the nature of the compact objects \cite{LIGOScientific:2018hze}. These constraints on the tidal effects allowed to set constraints on neutron star radii and on the EOS.
The radii of the binary components were estimated as to be $R_1=11.9^{+1.4}_{-1.4}$ km (heavier star) and $R_2=11.9_{-1.4}^{+1.4}$ km (lighter star) \cite{LIGOScientific:2018cki}.
Moreover, the electromagnetic
counterparts of GW170817 have set additional 
constraints on the lower limit of the tidal deformability: $\Lambda_{1.37M_\odot} > 210$ \cite{Bauswein2019}, 300 \cite{Radice2018},
279 \cite{Coughlin2018}, and 309 \cite{Wang2018}.

The expected increase in the number of detections of BNS coalescences in the upcoming observing runs of second-generation detectors~\cite{LVK-prospects} and future experiments~\cite{Maggiore:2020,Evans:2021,Hall:2022}, will most likely help tighten current constraints. Not surprisingly, the prospects to further improve the knowledge on the NS EOS using upcoming  tidal deformability information from inspiral BNS waveforms has received considerable attention~\cite{Delpozzo:2013,Lackey:2015,Abdelsalhin:2018,Hernandez-Vivanco:2019,Wang:2020,Landry:2020}. Existing approaches are based on Bayesian inference procedures and attempts considering alternatives routes, such as machine learning, are generally lacking, although steps have already been taken in that direction~\cite{Morawski:2020}. Here, this path is taken and an assessment of the performance of a machine learning model in classifying the NS EOS from the analysis of the tidal deformability parameter of a simulated dataset of inspiral GW signals from BNS coalescences, is made. Machine Learning (ML) is becoming a powerful tool in the analysis of GW data and the list of applications is rapidly growing \cite{Huerta:2019rtg,Cuoco:2020ogp}. 
Deep learning methods based on convolutional neural networks (CNN) have been used for binary black hole (BBH) signal detection and multiple-parameter estimation, reaching accuracies comparable to matched-filtering methods with higher computational efficiency~\cite{George:2016hay}. CNNs also show similar performance as matched-ﬁltering for BNS mergers but they are more resilient to non-stationary, non-Gaussian noise (glitches) in the GW data \cite{George:2017pmj}. Deep learning models have been used for instantaneous Bayesian posterior approximation in GW parameter estimation in \cite{Chua:2019wwt}. The use of masked autoregressive flows as a way of increasing the flexibility of deep neural networks in modeling GW posterior distributions was studied in \cite{Green:2020hst}. Detection of GW events with deep learning and their comparison with standard inference codes was carried out in \cite{Dax:2021tsq,Osvaldo:2021}, showing a fast and accurate inference of physical parameters. The development of fast ML alternatives to the computationally demanding Bayesian inference approaches has been remarkably established by \cite{Gabbard:2019rde} applying conditional variational autoencoders to BBH signals. ML approaches based on object detection have also  been applied successfully in the task of detecting BNS coalescence events from the GW data of current detectors~\cite{Aveiro:2022}.
Motivated by this growing body of work, here the purpose is to study the potential to infer the NS EOS directly from simulated BNS inspiral signals using a ML approach.

Across all fields of ML, convolutional-based networks (such as CNN~\cite{lecun1995convolutional}, arguably the most common architecture) have seen wide adoption. 
However, despite their success and popularity these networks have biases inherent to them (such as spatial locality and translation equivariance), which hinder their performance in certain tasks.  Conversely, models based on self-attention do not share these inductive bias~\cite{2021arXiv211009784G} and, additionally, they do not have non-degenerated convolutions~\cite{2021arXiv211106377H}. Due to this, they have seen increasing adoption in recent years and have been able to outperform convolutional networks in different tasks. For the Imagenet-1k Dataset~\cite{5206848}, for instance, the model with the best performance, PeCo~\cite{dong2021peco}, is based on the transformer architecute. The same is seen for the CIFAR-10 Dataset~\cite{krizhevsky2009learning}, where the model ViT-H/14~\cite{2020arXiv201011929D} shows the highest correct percentage.

In the present work the Audio Spectrogram Transformer (AST)  model~\cite{gong2021ast}, which is a purely attention-based and convolution-free network,  is applied to the analysis of GW data and its performance is explored. The goal is to use the AST model to classify  the NS EOS from a dataset of simulated inspiral signals of BNS mergers. 

The paper is organized as follows: the generation of all datasets used is described in Section \ref{sec:dataset} and the model and training procedure are presented in Section \ref{sec:model}. Results of the test applied to the trained models are discussed in Section \ref{sec:test} and a possible application of the model is explored in Section \ref{sec:additionalmodel}. Lastly, the conclusions are drawn in Section \ref{sec:conclusions}.\\

\section{Dataset\label{sec:dataset}}

\begin{figure}[!t]
    \includegraphics[width=\linewidth]{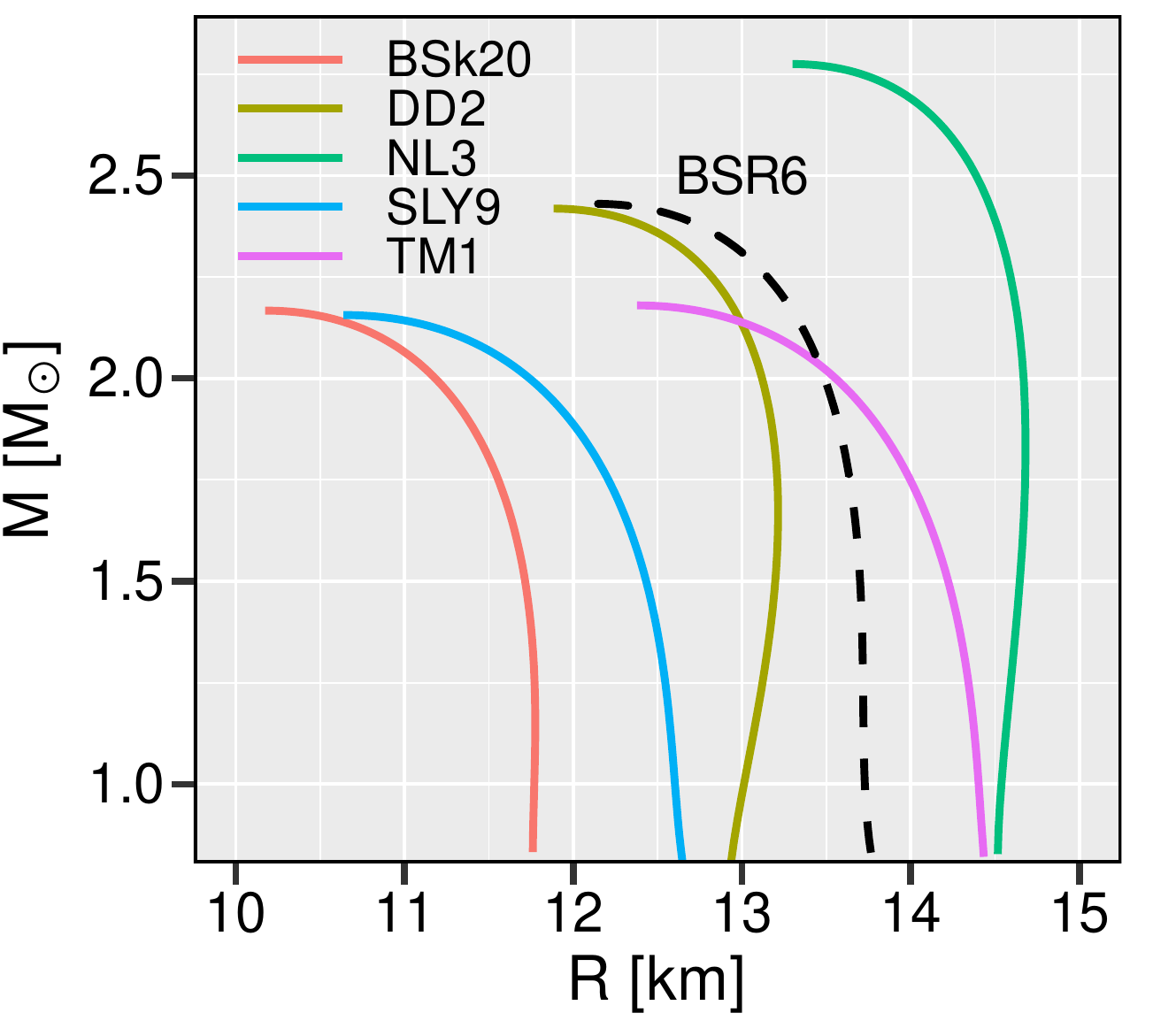}
    \caption{Mass-radius diagram for the set of EOS used in this work.}
    \centering
    \label{fig:mass_plot}
\end{figure}

\subsection{Neutron star matter EOS}

\begin{figure}[t]
     \includegraphics[width=\linewidth]{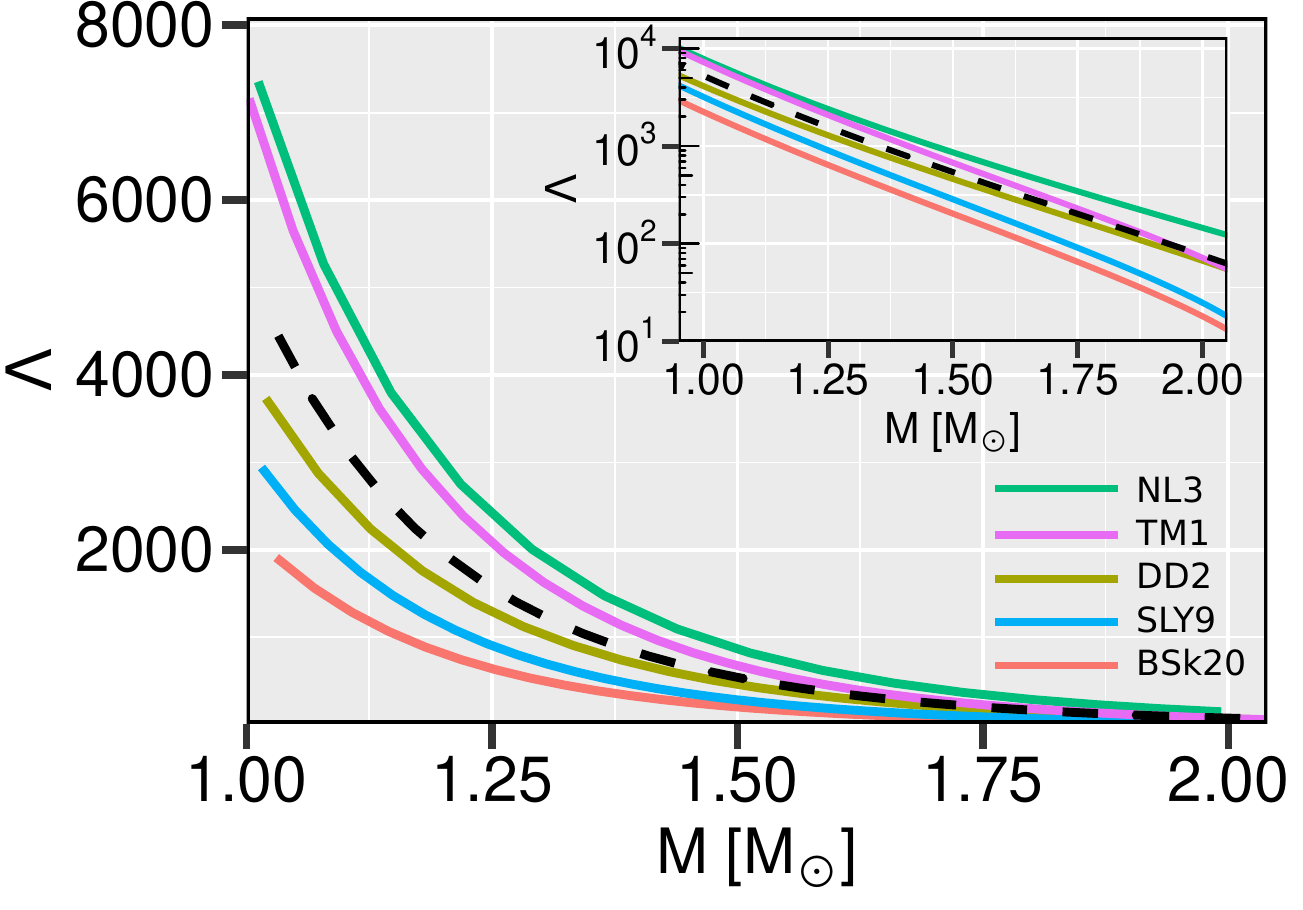}
    \caption{The diagram $\Lambda(M)$ for the set of EOS used in this work.
    The dashed line represent the BSR6 EOS that is explored in the last part of the work.}
    \centering
    \label{fig:mas_plot}
\end{figure}

For the AST model to be sensitive to a wide range of possible EOS in the mass-radius, $M(R)$, diagram, five representative EOS have been selected. These are plotted in solid lines in Fig.~\ref{fig:mass_plot} and they correspond to NS matter covering the ranges $2.16<M_{\text{max}}/M_{\odot}< 2.77$, $11.74<R_{1.4M_{\odot}}/\text{km}< 14.63$, and $11.18<R_{2.0M_{\odot}}/\text{km}< 14.67$.
All five EOS are unified EOS have been built from  either a relativistic mean field (RMF) approach or non-relativistic Skyrme interactions, and satisfy the constraint on the maximum NS mass $M_{\text{max}}\ge 2M_{\odot}$ (the details can be found in \cite{Fortin:2016hny}). The RMF EOS are divided into two types: i) nonlinear Walecka models with constant coupling, NL3 \cite{Lalazissis:1996rd} and TM1 \cite{Sugahara:1993wz}; and ii) density-dependent coupling parameters, DD2 \cite{Typel:2009sy}. Two Skyrme EOS are used: BSk20 \cite{Goriely:2010bm} and Sly9 \cite{chabanat1995interactions}.
The NS properties were determined by solving the
Tolmann-Oppenheimer-Volkoff (TOV) equations \cite{TOV1,TOV2}
together with the differential equations that determine
the tidal deformability \cite{Hinderer2008}. Figure \ref{fig:mass_plot} shows the $M(R)$ diagram and 
Fig.~\ref{fig:mas_plot} displays the tidal deformability dependence on the NS mass for each EOS. The NS tidal deformability is given by 
$\Lambda = \frac{2}{3}k_2C^{-5}$, where $k_2$ is the quadrupole tidal Love number and $C=M/R$ is the star's compactness \cite{Hinderer2008}. The properties of the GW emission from a BNS coalescence during the inspiral phase are sensitive to the EOS via the relations
$\Lambda_{\text{EOS}}(M)$. Figures~\ref{fig:mass_plot} and \ref{fig:mas_plot}, also show the BSR6 model \cite{Agrawal:2010wg} (dashed line), a nonlinear Walecka model with constant coupling, selected to test the ML model in the last section.

\subsection{Dataset generation}
\subsubsection{Waveform Generation}

All datasets used for training, validating and testing were obtained with the \texttt{PyCBC} package \cite{alex_nitz_2022_6324278}, employing the {\tt IMRPhenomPv2-NRTidalv2} waveform approximant \cite{PhysRevD.100.044003}. 
To generate the inspiral waveform of a compact binary, both the intrinsic parameters that describe the components of the binary and the extrinsic parameters that specify the location and orientation of the binary with respect to the observer are required. 
The intrinsic parameters are the component masses $M_1$ and $M_2$, their spins $S_1$ and $S_2$, and tidal deformabilities $\Lambda_1$ and $\Lambda_2$.

The sampling procedure consists in randomly choosing values for $M_{1,2}$ from a uniform distribution in the range $[1.0, 2.0]M_{\odot}$ and imposing $M_1 \geq M_2$ (adhering to the convention $M_1 \geq M_2$ for the binary mass components).
The corresponding tidal deformability parameters, $\Lambda_1$ and $\Lambda_2$, were obtained from the $\Lambda(M)$ curve for each EOS. 
The spins of the binary NS components were set to zero in all generated waveforms. The extrinsic parameters that specify the localization and orientation of the binary system were fixed as follows.
In all instances, the inclination of the binary was set to zero and its distance was set to 39 Mpc. 
After generating the GW strain data and normalizing it to its maximum amplitude (guaranteeing that the amplitude of the strain is in the range $[-1,1]$ at all points of the time series), a cut is applied to the data so that 8092 points are kept around the peak of the strain (at merger). The resulting array is padded, with zeros, to a final length of 8192 points. Since the sample rate of the generated strain is 4092~Hz, this processing results in a time window of 2 s that encloses the merger.
This work uses three datasets that were constructed in the following way. First, two datasets were generated, Train set 3 (containing 750000 waveforms) and Test set 3 (containing 10000 waveforms), each with the constraint $M_{1,2}/M_{\odot} \in [1.0,2.0]$.
Then, from each of the two previous datasets, two subsets were filtered satisfying  $M_{1,2}/M_{\odot} \in [1.0,1.5]$ (Train set 1 and Test set 1) and $M_{1,2}/M_{\odot} \in [1.5,2.0]$ (Train set 2 and Test set 2). A similar approach was employed for the third dataset, used in Sec.~\ref{sec:additionalmodel}, where a dataset (BSR6 set 3) was generated with 1000 samples, followed by the creation of two subsets (BSR6 set 1 and 2).
Detailed information about the datasets can be found in Table \ref{table:datasets}, with the respective range of the binary properties. 

It should be noted that this study, as a first step, employs noise-free datasets, i.e.~no injections in noise, either Gaussian or real noise from the LIGO-Virgo detectors, are performed.

\setlength{\tabcolsep}{4pt} 
\renewcommand{\arraystretch}{1.8} 
\begin{table*}
\begin{center} 
\caption{All datasets generated and the respective ranges of some NS and binary properties.
$M_{1,2}$ and $\Lambda_{1,2}$ are the mass and tidal deformability of each NS binary component, respectively.
The range for the binary chirp mass is also reported,
${\cal M}_{\text{chirp}}=(M_1M_2)^{3/5}/(M_1+M_2)^{1/5}$,
the effective tidal deformability of the binary,
$
\tilde{\Lambda}=(16/13)((M_1+12M_2)M_1^4\Lambda_1+(M_2+12M_1)M_2^4\Lambda_2)/(M_1+M_2)$, and the binary mass ratio, $q=M_2/M_1$.}

\begin{tabular}{c c c c c c c c}
 \hline
  Models & Train Datasets & Samples [\#] & $M_{1,2} [M_{\odot}]$ & $\Lambda_{1,2}$  & $\tilde{\Lambda}$ & q & ${\cal M}_{\text{chirp}} [M_{\odot}]$ \\ [0.5ex] 
 
 \hline 
 \makecell{Model A \\ Model B \\ Model C} &
 \makecell{ Train Set 1 \\ Train Set 2 \\ Train Set 3 } & 
 \makecell{ 187 811 \\ 186 718 \\ 750 000 } &
 \makecell{ $[1.0,1.5]$ \\ $[1.5,2.0]$ \\ $[1.0,2.0]$  } &
 \makecell{ $[203,7902]$ \\ $[18, 869]$ \\ $[18,7902]$ } &
 \makecell{ $[206,10157]$ \\ $[18, 936]$ \\ $[18,16137]$ } &
 \makecell{ $[0.67,1]$ \\ $[0.75,1]$ \\ $[0.50,1]$ } &
 \makecell{ $[0.87,1.30]$ \\ $[1.31,1.74]$ \\ $[0.87,1.74]$ } \\
 \hline
 \\
  \hline
  &Test Datasets & Samples [\#] & $M_{1,2} [M_{\odot}]$ & $\Lambda_{1,2}$  & $\tilde{\Lambda}$ & q & ${\cal M}_{\text{chirp}} [M_{\odot}]$ \\ [0.5ex] 
 
 \hline 
 \makecell{} &
 \makecell{ Test Set 1 \\ Test Set 2 \\Test Set 3 } & 
 \makecell{ 2 491 \\ 2 496 \\ 10 000 } &
 \makecell{ $[1.0,1.5]$ \\ $[1.5,2.0]$ \\ $[1.0,2.0]$ } &
 \makecell{ $[204,7901]$ \\ $[18,866]$ \\ $[18, 7901]$ } &
 \makecell{ $[214,9592]$ \\ $[19,894]$ \\ $[19, 14616]$ } &
 \makecell{ $[0.67,1]$ \\ $[0.76,1]$ \\ $[0.50,1]$ } &
 \makecell{ $[0.87,1.30]$ \\ $[1.31,1.74]$ \\ $[0.88,1.74]$ } \\
 \hline \\
 
  \hline
  &BSR6 Datasets & Samples [\#] & $M_{1,2} [M_{\odot}]$ & $\Lambda_{1,2}$  & $\tilde{\Lambda}$ & q & ${\cal M}_{\text{chirp}} [M_{\odot}]$ \\ [0.5ex] 
 
 \hline 
 \makecell{} &
 \makecell{ BSR6 Set 1 \\ BSR6 Set 2 \\ BSR6 Set 3 } & 
 \makecell{ 251 \\ 246 \\ 1 000 } &
 \makecell{ $[1,1.5]$ \\ $[1.5,2.0]$  \\ $[1.0,2.0]$ } &
 \makecell{ $[542,5229]$ \\ $[76,540]$ \\ $[76, 5231]$ } &
 \makecell{ $[587,5929]$ \\ $[81,543]$ \\ $[81,9758]$ } &
 \makecell{ $[0.68,1]$ \\ $[0.75,1]$ \\ $[0.51,1]$ } &
 \makecell{ $[0.89,1.29]$ \\ $[1.31,1.73]$ \\ $[0.89,1.72]$ } \\
 \hline
 
 \label{table:datasets} 
\end{tabular}

\end{center}
\end{table*}


\subsubsection{Spectrogram Generation}

\begin{figure}[t!]
    \centering
    \includegraphics[width=60mm]{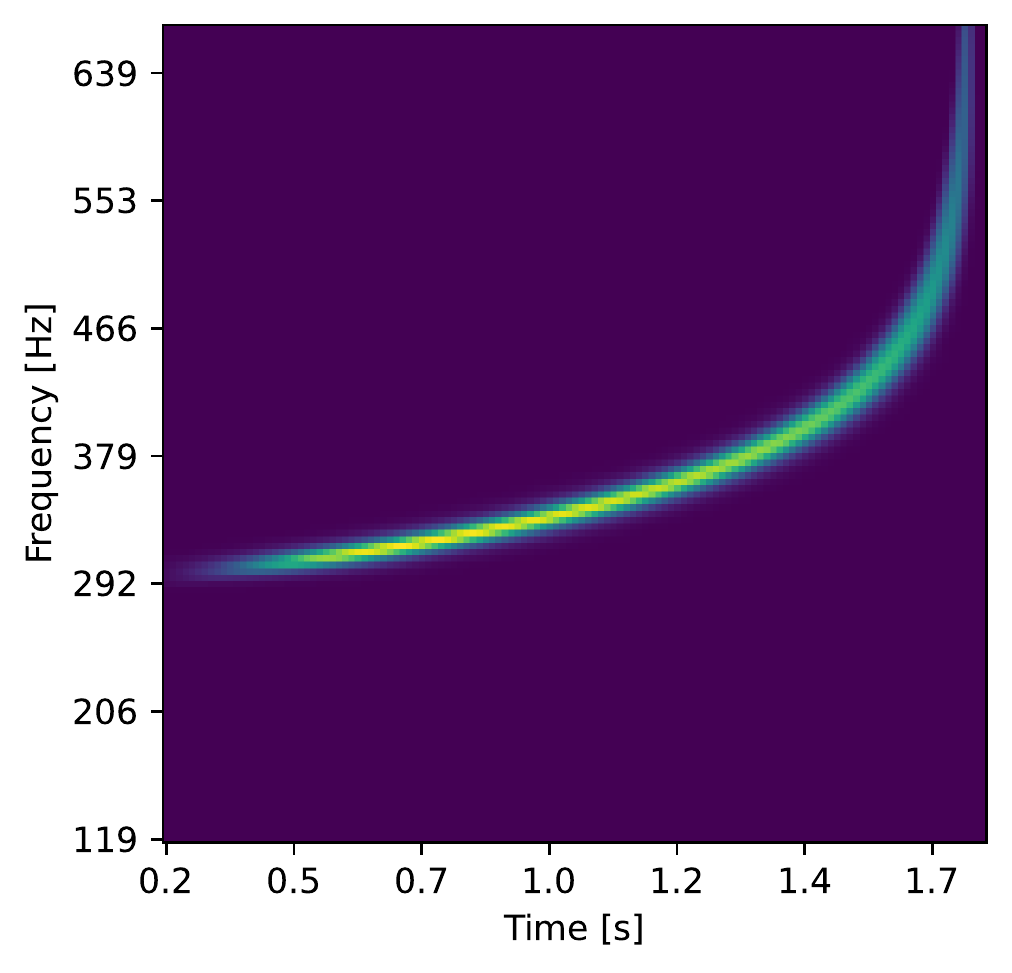} 
    \caption{Example of spectrogram from the dataset.}
    \label{fig:spec_ex}
\end{figure}

Having the waveform in the desired shape, all that is left is converting it to a spectrogram. For this purpose, the Constant-Q Transform \cite{schorkhuber_2010} supplied by the {\tt librosa} package \cite{brian_mcfee_2022_6097378} is used. For this transform, the following parameters were employed: $128$ frequency bins, a minimum frequency of $\SI{32}{Hz}$, a length of $64$, $28$ bins per octave and a tuning parameter set to zero. The resulting spectrogram has 128 frequency bins and 129 time bins. Figure~\ref{fig:spec_ex} shows one example of a generated spectrogram.

\section{Model \label{sec:model}}

The AST model takes the input spectrogram, splits it into several $16\times 16$ patches and proceeds to flatten each patch to a 1D patch embedding of size 768, using a linear projection layer. An additional trainable positional embedding is added to each of the previous embeddings and the resulting sequence is used as input for the Transformer \cite{vaswani2017attention}. This allows for the option of transfer-learning with other models that make use of the standard Transformer architecture.

Although the AST model can be trained from scratch, it has shown increased performance when the model is initialized with weights originated from previously trained models. Due to this, the weights from a Data-efficient Image Transformer (DeiT) \cite{touvron2021training}, trained on the ImageNet dataset \cite{5206848}, were used. This model - referred to as DeiT-base distilled 384 - was chosen because it showed very good Top-1 Accuracy\footnote{This metric is calculated by dividing the total number of correct predictions, $p_{c}$, by the total number of predictions, $p_{t}$, and multiplying by 100, or $p_{c}$/$p_{t}$ x 100.} with fewer parameters, when compared to the state of the art model PeCo \cite{dong2021peco}, as is presented in Table~\ref{table:pecovsdeit}.  

\begin{center}
\begin{table}[t]
\caption{Number of parameters and Top-1 Accuracy (achieved on the ImageNet-1k dataset \cite{5206848}) comparison of the model used for pre-training, DeiT-base distilled 384 \cite{touvron2021training}, and the stateof the art model PeCo \cite{dong2021peco}.}
\begin{tabular}{c c c}
 \label{table:pecovsdeit}
 \\
 \hline
 Model Name & \# Params & Top-1 Accuracy \\ [0.5ex] 
 \hline
 DeiT-base distilled 384 & 87M & 85.2\% \\ 
 
 PeCo & 656M & 88.3\%  \\
 \hline
 
\end{tabular}

\end{table}
\end{center}

\subsection{Training Configuration\label{sec:model_train_config}}
This section gives an overview of the training process for the models.

\subsubsection{Hardware and Resources}
A NVIDIA A100 Tensor Core GPU with 40GB of memory was used for training, which allowed for a batch size of 480 samples. The training was carried out for 50 epochs which took around 11 hours and 40 minutes (approximately 14 min/epoch).

\subsubsection{Learning Rate and Loss Function}
The optimizer used for training was Adam \cite{kingma2014adam}. The parameters used for this optimizer were set as follows: the starting learning rate to $\text{lr}=\mathrm{10}^{-5}$, the weight decay to $\alpha=5\times\mathrm{10}^{-7}$, and $\left( \beta_{1}, \beta_{2} \right) = \left( 0.95, 0.999 \right)$. The weight decay parameter controls the regularization amount and the $\beta_{1,2}$ control the running averages of gradient and its square on the Adam algorithm. 
The scheduler MultiStepLR was also employed to reduce the learning rate by half at epochs 5, 10, 20 and 30.

The Loss Function used was BCEWithLogitsLoss (which is the standard Cross Entropy Loss inside a sigmoid function).

\subsubsection{Data Augmentation}
The purpose of Data Augmentation is to leverage suitable data transformations to achieve an increase in flexibility, variability and generalization ability of the model. In the case of image classification, many transformations commonly used (such as rotation, translation, resizing, flipping, ...) were designed to help classify ordinary objects like fruits or vehicles, not spectrograms. Due to this, in this work only two methods of data augmentation were employed, SpecAugment \cite{park2019specaugment} and mixup \cite{zhang2017mixup}.

\begin{figure}[t!]
    \centering
    \includegraphics[width=\linewidth]{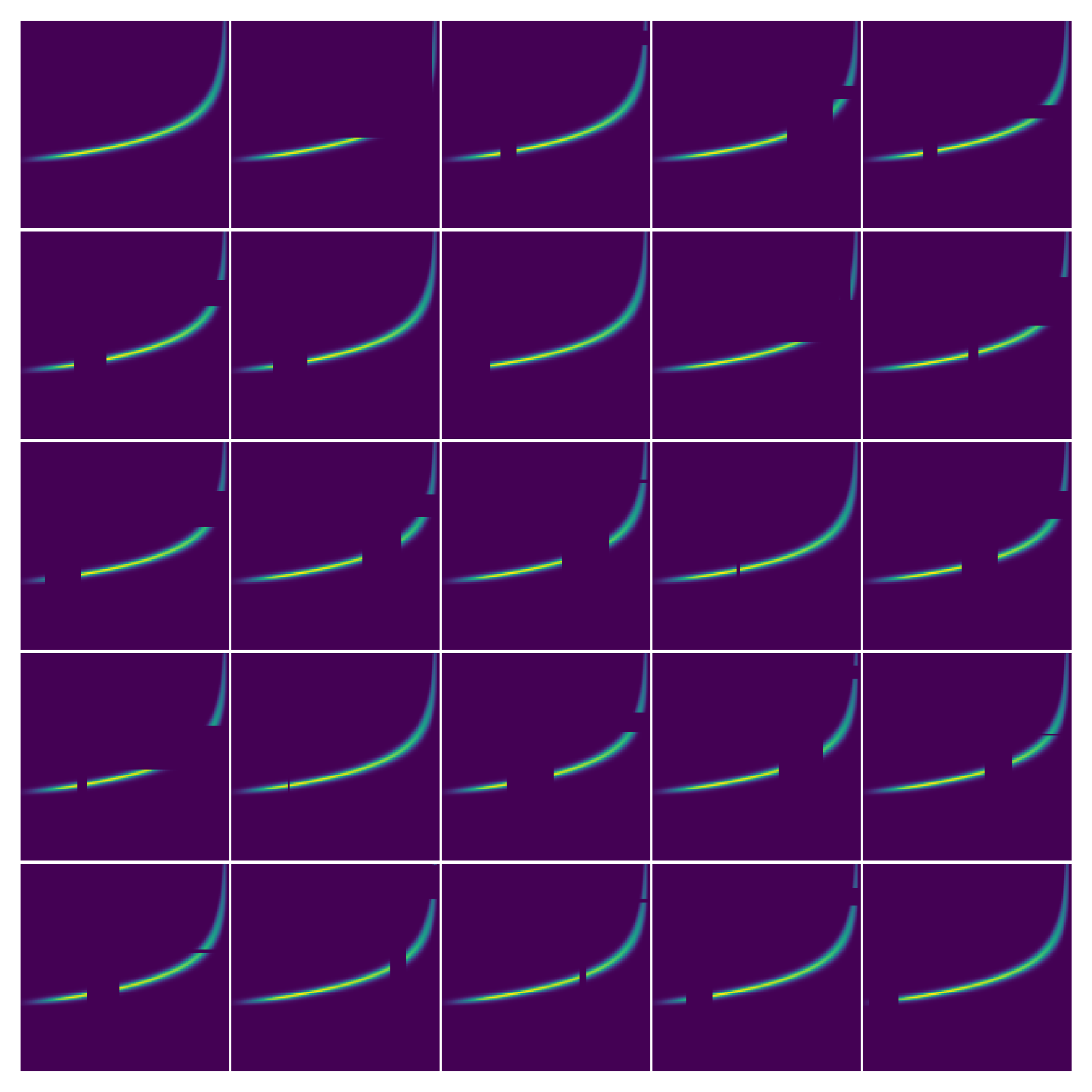} 
    \caption{SpecAugment method. The top left image corresponds to the original, un-augmented image and the following 24 correspond to possible augmentations.}
    \label{fig:specaug}
\end{figure}

The SpecAugment method consists in applying masking\footnote{The term "masking" means to remove/omit the values of certain bins from the spectrogram.} to the spectrogram, in both the frequency and time domains. 
Frequency masking is applied so that \textit{f} consecutive bins $[f_0,f_0+f)$ are masked, where \textit{f} is sampled from a uniform distribution from 0 to the frequency mask parameter \textit{F}. The initial frequency \textit{$f_0$} is chosen from $[0,\nu-f)$ where $\nu$ is the total number of frequency bins.
Similarly, time masking is applied so that \textit{t} consecutive time steps $[t_0,t_0+t)$ are masked, where t is sampled from an uniform distribution from 0 to the time mask parameter \textit{T}. The initial time $t_0$ is chosen from $[0,\tau-t)$, where $\tau$ is the total number of time bins. Since the number of steps masked is re-calculated at every epoch, it was found that setting  the mask parameters to $\textit{F} = \frac{\nu}{2}$ and $\textit{T} = \frac{\tau}{2}$, allowed for a high degree of variability while still preserving sufficient information in the resulting spectrogram, as can be seen in Figure~\ref{fig:specaug}.

\begin{figure}[t!]
    \centering
    \includegraphics[width=\linewidth]{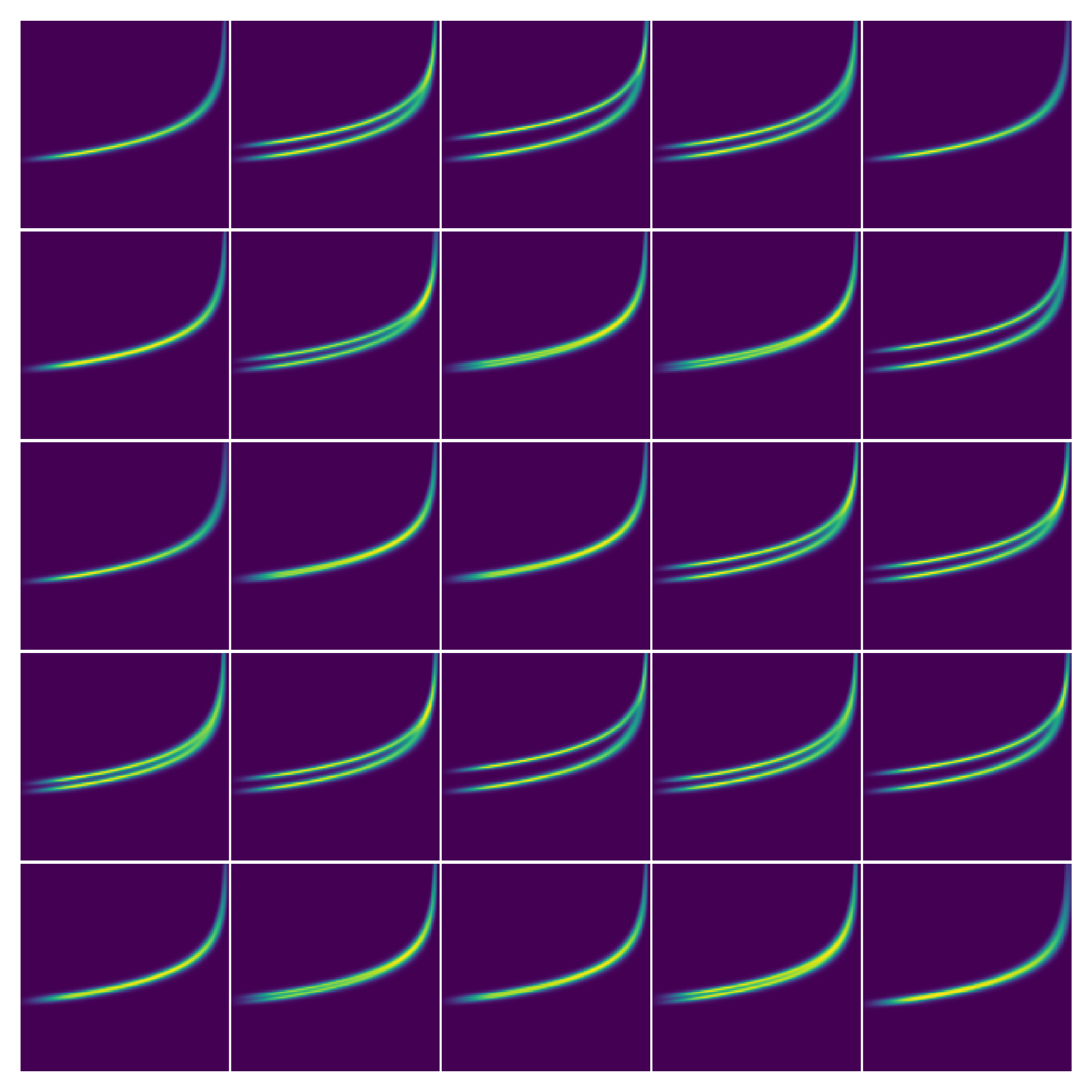} 
    \caption{Mixup method. The top left image corresponds to the original, un-augmented image and the following 24 correspond to possible augmentations.}
    \label{fig:mixup}
\end{figure}

The mixup is a simple data-agnostic data augmentation routine that constructs virtual training samples as $\tilde{x} = \lambda x_i + (1-\lambda)x_j$ and $\tilde{y} = \lambda y_i + (1-\lambda)y_j$, where $x_i$ and $x_j$ are two distinct spectrograms and $y_i$ and $y_j$ are their respective one-hot labels. So, from two samples, $(x_i,y_i)$ and $(x_j,y_j)$, a virtual sample is created $(\tilde{x},\tilde{y})$ and passed on to the model. For this method, $\lambda$ was sampled from a $\beta$ distribution, using Numpy's function \texttt{numpy.random.beta()} with $\alpha = \beta = 10$. The impact of this sampling can be seen in Figure~\ref{fig:mixup}. This augmentation method had a crucial effect in stabilising the learning process and on the model's generalisation, by suppressing oscillations when predicting outside the training set.

\section{Tests\label{sec:test}}

In this section, the success of three models independently trained is explored. It should be noted that the same configuration (detailed in Section \ref{sec:model_train_config}) was used for all models, and the only difference comes from the dataset used for training (Table \ref{table:datasets}).

Firstly, Model A, with each star's mass limited to $[1.0,1.5]M_{\odot}$ (train set 1) was trained. The reason for this mass range is that this is the region where the EOS curves used as benchmarks in this study are mostly distinguishable from one another (cf.~Fig.~\ref{fig:mas_plot}). Hence, it is the region where the model should most easily converge. To contrast with the first model, a second model (Model B) was trained on $M_{1,2}/M_{\odot} \in [1.5,2.0]$ (train set 2). Additionally, a third model, Model C,  was also trained but on the full parameter range $[1.0,2.0]M_{\odot}$ (train set 3). The goal is to compare the performances between full and shorter range models and define the best approach when dealing with detections. After the training phase, each model is tested on new datasets, generated exclusively for testing, with the mass constraints of the stars equal to the ones used for the respective model's training. Additionally, model C is also tested on test sets 1 and 2, in order to enable a performance comparison with the other two models.

To evaluate the performance of the model in the test datasets, a confusion matrix together with a table are presented, for each case. Along the horizontal lines, each number in the confusion matrix   represents the number of samples predicted with the corresponding class. The metrics Distance Score, Top-1\footnote{This metric can also be obtained from the confusion matrix. To do this, choose a EOS and divide the value of the corresponding  diagonal square by the sum of all the values in the squares of the matching horizontal line.} and Top-2 Accuracy, reported on the tables, are described as follows:
\\\\
\textbf{Top-n Accuracy:} Given a sample, the model outputs a vector, \textit{$\overrightarrow{v}$}, with size equal to the number of classes. In this vector each element \textit{$v_{i}$} is interpreted as the predicted probability that the given sample belongs to the \textit{i-th} class. The top-n accuracy metric expresses the average percentage of instances where the target class was within the \textit{n} predicted classes with the highest probabilities. \\

\textbf{Distance Score:} This metric, calculated by averaging across samples, gives the difference between the predicted probability that a given sample belongs to the correct class  (i.e. the label) \textit{$p_{\rm label}$} and the predicted probability that the same sample belongs to a given class, \textit{$p_{\rm class}$}.
\begin{equation}
    \label{eq:ds}
    ds = p_{\rm label} - p_{\rm class}\,.
\end{equation}
As an example assume, for a given sample, the label is BSk20 and it is interesting to know the distance score of EOS SLy9. If, in this case, the \textit{$p_{\rm BSk20}$}~=~60\% and \textit{$p_{\rm SLy9}$}~=~20\% then the \textit{ds} is 40\%, according to Eq.~(\ref{eq:ds}).

Although this metric might seem simple, it provides very useful information. In the extremes, $ds=0$ means that, on average, the predicted probability that the given sample belongs to the correct class is equal to another class, which can be interpreted as the model being very indecisive between at least those two classes (this behaviour is expected when the sample belongs to a region with big overlap between the two corresponding EOS). Conversely, $ds=100$\% means that, on average, the probability that a sample belongs to the correct class was predicted as 100\%, which implies that the model is able to distinguish between classes with complete certainty. On the other side, $ds=-100$\% indicates that the model completely failed its predictions.

In this work, this metric will only be calculated when the model misclassifies a sample and the class chosen for the calculation will be the one predicted with the highest probability. Due to this, the value of \textit{ds} can be interpreted as how close - percentage-wise - the model is to making the correct prediction. The consequence of this choice is that \textit{ds} $\in [-100,0]$ and the closer its value is to 0, the better. 

\subsection{Model A ($M_{1,2}/M_{\odot} \in [1.0,1.5]$)}
\label{model_1_15}

\begin{figure}[t!]
    \centering
    \includegraphics[width=50mm]{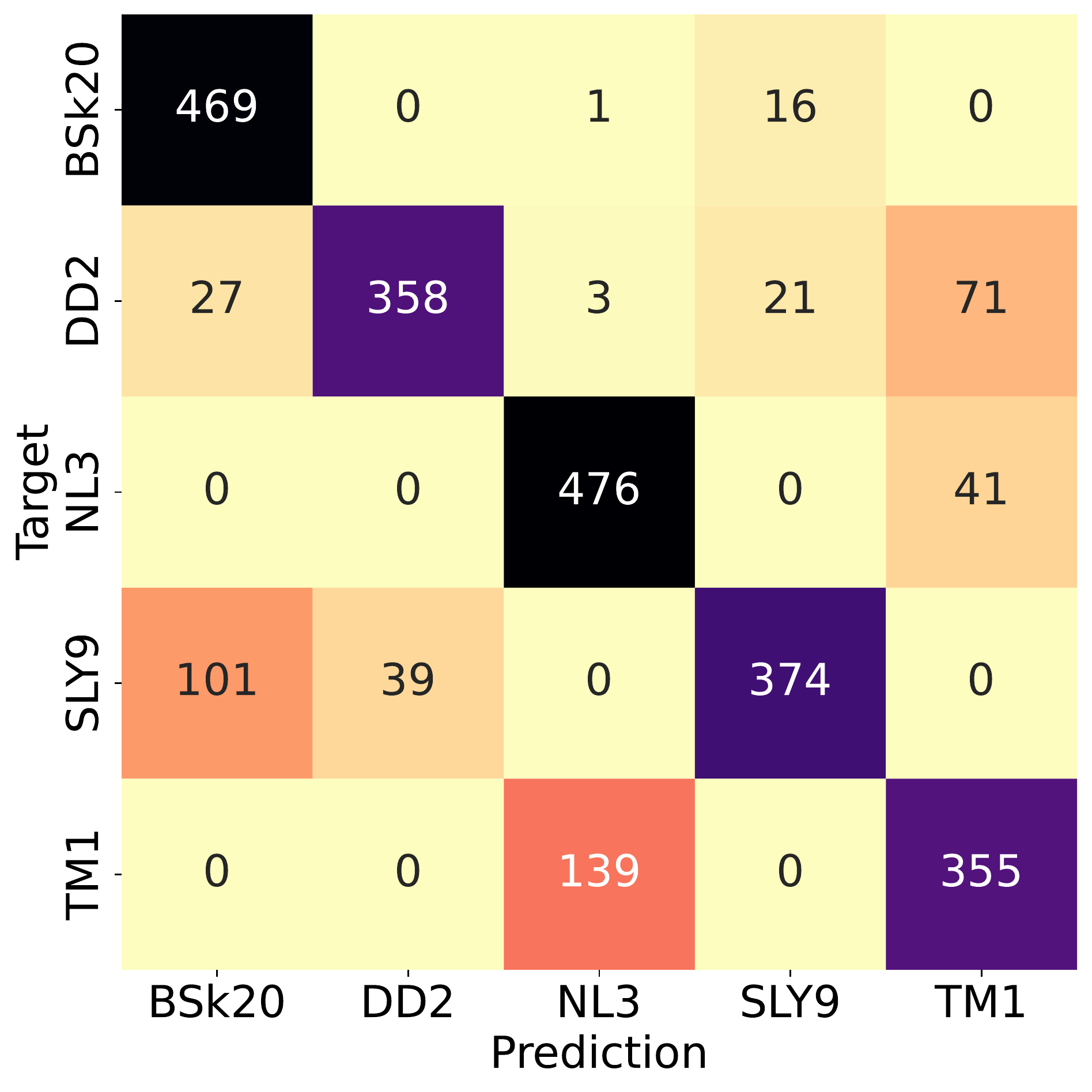} 
    \caption{Confusion matrix obtained for Model A (see Table \ref{table:datasets}) on test set 1, $M_{1,2}/M_{\odot} \in [1.0,1.5]$.}
    \label{fig:m_1_15}
\end{figure}

The performance of the model on the train set 1, $M_{1,2}/M_{\odot} \in [1.0,1.5]$, is reported in Table~\ref{table:distance_1_15}. It is found that the model best classifies samples from BSk20 with a top-1 accuracy of 96.5\%. On the other hand, the model has the worst performance for samples from TM1, but still reaches a top-1 accuracy of 71.9\% . While there is, approximately, a 25\% difference in these scores, values above 70\% are already promising.

\setlength{\tabcolsep}{4pt} 
\renewcommand{\arraystretch}{1.8} 
\begin{table}[t!]
\begin{center} 
\caption{Results for the top-1 and top-2 accuracy and for the distance score $ds$ obtained by Model A (see Table \ref{table:datasets}) on the test set 1, $M_{1,2}/M_{\odot} \in [1,1.5]$. The last column indicates the average values of the metrics for all EOS.}
\begin{tabular}{c c c c c c c}
 \hline
  metric [\%] & BSk20 & DD2 & NL3 & SLy9 & TM1 & \textbf{Avg} \\ [0.5ex] 
 
 \hline
 \makecell{top-1 \\  top-2 \\ \textit{ds}} & 
 \makecell{96.5 \\ 99.4 \\ -7.39} &
 \makecell{74.6 \\ 81.5 \\ -8.89} &
 \makecell{92.1 \\ 100.0 \\ -9.06} &
 \makecell{72.8 \\ 99.6 \\ -15.37} &
 \makecell{71.9 \\ 99.2 \\ -13.16} &
 \makecell{\bf{81.58} \\ \bf{95.94} \\ \bf{-10.77}} \\

 \hline
 
 \label{table:distance_1_15} 
\end{tabular}

\end{center}
\end{table}

Although the model is  unlikely to misclassify a sample from the classes corresponding to the upper and lower bounds of the $\Lambda(M)$ curves (i.e.~BSk20 and NL3), when it does  \textit{ds} is higher than -10\%. More interestingly, even for the classes in between, where misclassification is more common, it only decreases to, at most, -15.37\% indicating that, even in failure, the model was close to making the correct prediction. This is corroborated when looking at the top-2 accuracy where, for the DD2 class, it has a value of 81.5\%, which is already quite good, but is surpassed by the excellent values (greater than 99\%) seen in the other classes. 

\begin{figure}[t!]
    \includegraphics[width=\linewidth]{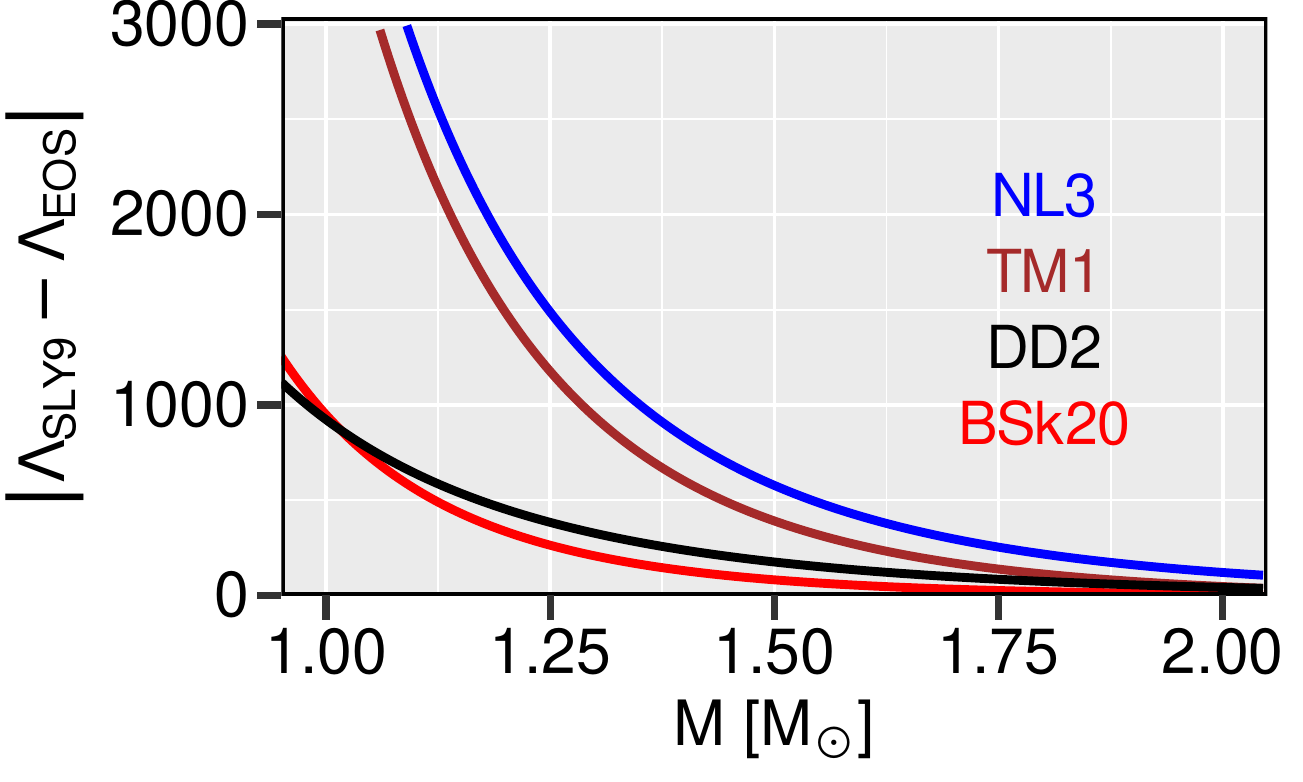}
    \caption{The prediction differences for the tidal deformability $\Lambda(M)$ between the SLy9 EOS and the remaining EOS of the samples.}
    \centering
    \label{fig:NL3_dif_plot}
\end{figure}

The results in Table~\ref{table:distance_1_15} indicate that the model's certainty appears to increase with the uniqueness of the class, i.e. with the unique behavior of the model that defines the class. The model misclassification pattern is shown in the confusion matrix displayed in Fig.~\ref{fig:m_1_15}. To understand this pattern Fig.~\ref{fig:NL3_dif_plot} shows, as an illustrative example, 
the difference for $\Lambda(M)$ between the SLy9 EOS and all other four EOS. It can be seen that the most similar EOS  to SLy9 (in terms of $\Lambda$) are both DD2 and BSk20. This fact explains why the model missclassifications of SLy9 were attributed to BSk20 (with 16 mismatches) and DD2 (with 21 mismatches). A similar argument explains the misclassifications shown for the other EOS in Figure~\ref{fig:m_1_15}.

 \subsection{Model B ($M_{1,2}/M_{\odot} \in [1.5,2.0]$).}\label{model_15_2}

\begin{figure}[t!]
    \centering
    \includegraphics[width=50mm]{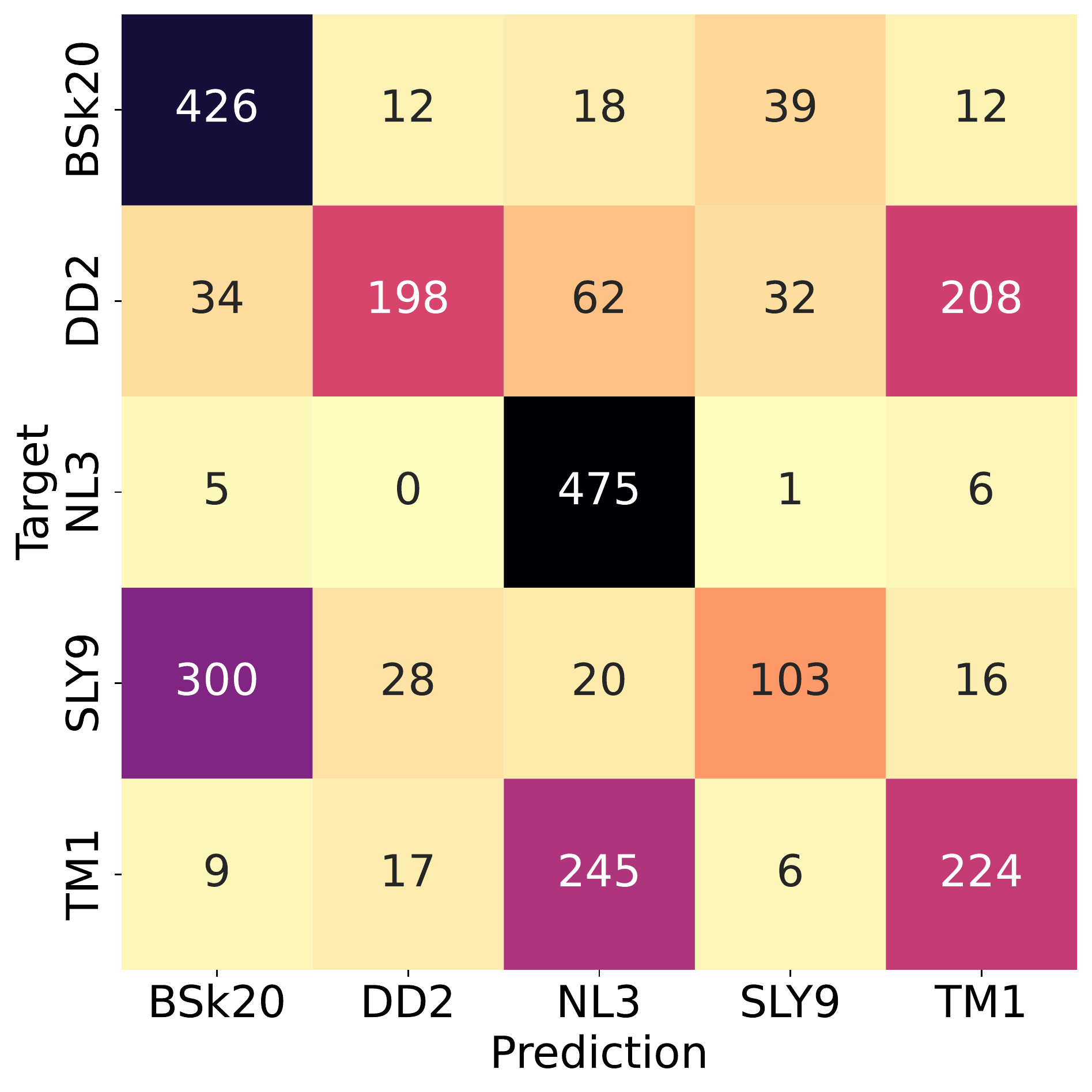} 
    \caption{Confusion matrix obtained for Model B (see Table \ref{table:datasets}) on the test set 2, $M_{1,2}/M_{\odot} \in [1.5,2.0]$.}
    \label{fig:m_15_2}
\end{figure}

Since this range corresponds to the region with the strongest overlap between classes (cf.~Fig.~\ref{fig:mas_plot}), a high degree of confusion is expected. This is shown in Figure~\ref{fig:m_15_2}. Although the model is able to classify well samples from BSk20 and NL3 (which, unsurprisingly, correspond to the upper and lower bounds for the $\Lambda(M)$ curves shown in Fig.~\ref{fig:mas_plot}), it is unable to do so for the classes in-between which, as previously mentioned, is anticipated in this mass range. Looking at Table~\ref{table:distance_15_2}, remarkably the model surpasses 90\% top-2 accuracy on all classes but DD2, where it reaches 73.6\%. The high values for the top-2 accuracy together with the high values of \textit{ds} (around -10\%), validate the hypothesis that the model's ability to classify the sample is correlated to the uniqueness of the sample's EOS. Even when the model is trained on a harsher  region (when compared to the $[1.0,1.5]M_{\odot}$ case) and looking only at the least 'unique' class, DD2, it was found that although the model misclassifies samples in 62.9\% of the cases, it is close to making the correct prediction (which is supported by the high top-2 accuracy of 73.6\% and the high \textit{ds} of -8.52\%). A similar discussion is applicable to the model's performance across all classes, since the model achieves, on average, a top-2 accuracy of 90.18\% and a \textit{ds} of $-10.29\%$, see last column of Table \ref{table:distance_15_2}.

\setlength{\tabcolsep}{4pt} 
\renewcommand{\arraystretch}{1.8} 
\begin{table}[t!]
\begin{center} 
\caption{Results for the top-1 and top-2 accuracy and for the distance score $ds$ obtained by Model B (see Table \ref{table:datasets}) on the test set 2, $M_{1,2}/M_{\odot} \in [1.5,2]$. The last column indicates the average values of the metrics for all EOS.}

\begin{tabular}{ c c c c c c c}
 \hline
  metric [\%] & BSk20 & DD2 & NL3 & SLy9 & TM1 & \textbf{Avg}  \\ [0.5ex] 
 
 \hline 
 \makecell{top-1 \\  top-2 \\ \textit{ds}} & 
 \makecell{84.0 \\ 91.7 \\ -9.54} &
 \makecell{37.1 \\ 73.6 \\ -8.52} &
 \makecell{97.5 \\ 98.8 \\ -6.96} &
 \makecell{22.1 \\ 90.6 \\ -12.06} &
 \makecell{44.7 \\ 96.2 \\ -14.38} &
 \makecell{\bf{57.08} \\ \bf{90.18} \\ \bf{-10.29}}  \\

 \hline
 
 \label{table:distance_15_2} 
\end{tabular}

\end{center}
\end{table}

 
\subsection{Model C ($M_{1,2}/M_{\odot} \in [1.0,2.0]$).\label{model_1_2}}

\begin{figure}[t!]
    \centering
    \includegraphics[width=50mm]{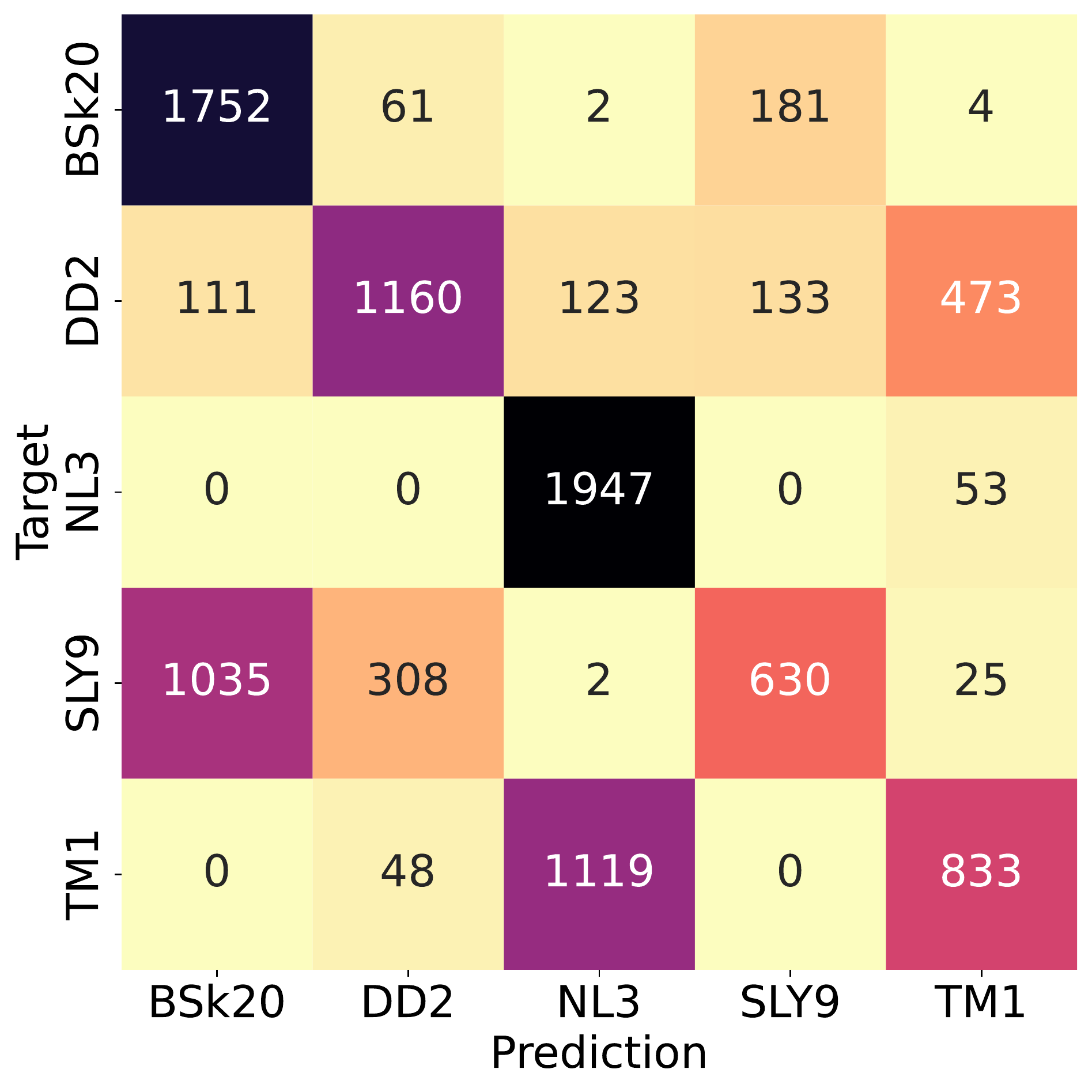}
        \caption{Confusion matrix obtained for Model C (see Table \ref{table:datasets}) on test set 3, $M_{1,2}/M_{\odot} \in [1.0,2.0]$.}
\end{figure}

For the case of the full mass range, the results presented in Table~\ref{table:distance_1_2} show that the top-1 accuracy from either BSk20 or NL3 are high (87.6\% and 97.4\%, respectively). The performance for samples from the other EOS classes is not as impressive, only reaching a top-1 accuracy of 43.7\% (averaged across the values seen for DD2, SLy9 and TM1). Since this range includes regions with overlap between classes, the values achieved for the top-2 accuracy should also be closely analysed. This  metric exhibits promising results reaching over 95\% on all classes but DD2, where 78.2\% is obtained. These results, together with the high values of \textit{ds} are encouraging and the success of the model is highlighted when observing its average behaviour in this range, where it accomplishes a top-1 accuracy of 63.22\%, top-2 accuracy of 93.64\% and \textit{ds} of -8.22\%, last column of Table~\ref{table:distance_1_2}.

\setlength{\tabcolsep}{4pt} 
\renewcommand{\arraystretch}{1.8} 
\begin{table}[t!]
\begin{center} 
\caption{Results for the top-1 and top-2 accuracy and for the distance score $ds$ obtained by Model C (see Table \ref{table:datasets}) on the test set 3, $M_{1,2}/M_{\odot} \in [1.0,2.0]$. The last column indicates the average values of the metrics for all EOS.}
%

\begin{tabular}{c c c c c c c}
 \hline
 metric [\%] & BSk20 & DD2 & NL3 & SLy9 & TM1 & \textbf{Avg} \\ [0.5ex] 

 \hline
 \makecell{top-1 \\  top-2 \\ \textit{ds}} & 
 \makecell{87.6 \\ 95.8 \\ -4.39} &
 \makecell{58.0 \\ 78.2 \\ -5.88} &
 \makecell{97.4 \\ 100 \\ -4.44} &
 \makecell{31.5 \\ 95.6 \\ -11.09} &
 \makecell{41.6 \\ 98.6 \\ -15.32} &
 \makecell{\bf{63.22} \\ \bf{93.64} \\ \bf{-8.22}} \\

 \hline
 
 \label{table:distance_1_2} 
\end{tabular}

\end{center}
\end{table}

Since this model was trained on the full mass range, it is expected to have a good performance across all the test datasets used. To enable a fair comparison between this model and the ones discussed in sections \ref{model_1_15} (Model A) and \ref{model_15_2} (Model B), it was additionally tested on the datasets used in those sections (test sets 1 and 2).
It is worth noting that the average of the values of the metrics, on both sets, is not expected to coincide with those reported in Table~\ref{table:distance_1_2}.  Since the mass restriction is applied to each individual star, samples corresponding to systems with, for example, $M_1/M_{\odot}=1.1$ and $M_2/M_{\odot}=1.8$ will not be present in either test set 1 or 2, due to one of the stars being outside the range, but will be included in test set 3 since the associated constraint is $M_{1,2}/M_{\odot} \in [1.0, 2.0]$. Hence, the test set 3 contains samples which are not available in either of the two other test sets.

\begin{figure}[t!]
    \centering
    \includegraphics[width=\linewidth]{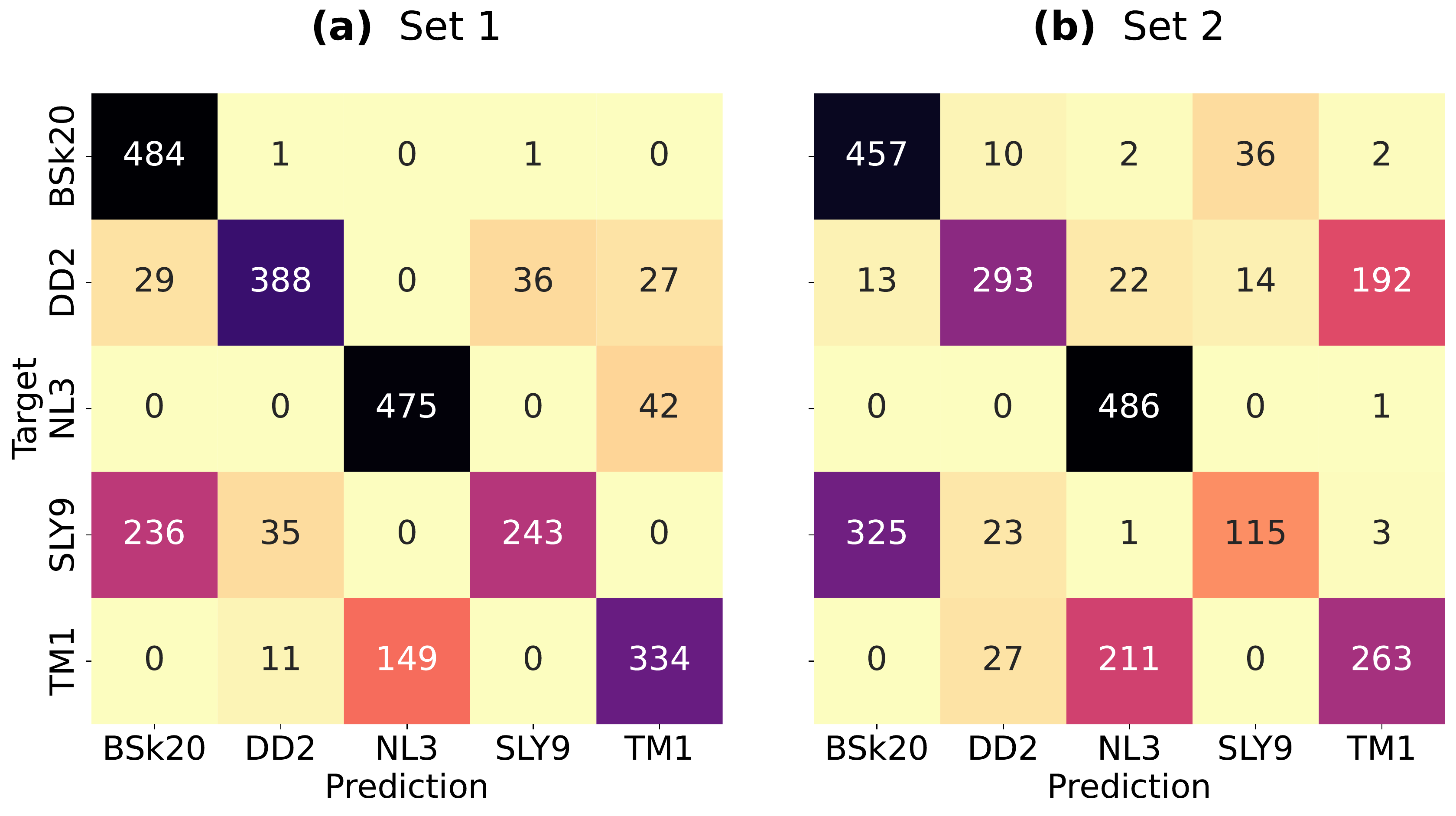} 
    \caption{Confusion matrices obtained for Model C on the test set 1 (left), $M_{1,2}/M_{\odot} \in [1,1.5]$, and on the test set 2 (right), $M_{1,2}/M_{\odot} \in [1.5,2]$.}
    \label{fig:m_1_2_other}
\end{figure}

\setlength{\tabcolsep}{2.5pt} 
\renewcommand{\arraystretch}{1.8} 
\begin{table}[t!]
\begin{center}
\caption{Results for the top-1 and top-2 accuracy and for the distance score $ds$ obtained by Model C (see Table \ref{table:datasets}) on the test set 1, $M_{1,2}/M_{\odot} \in [1,1.5]$ (top entrances), and on the test set 2, $M_{1,2}/M_{\odot} \in [1.5,2]$ (bottom entrances). The last column indicates the average values of the metrics for all EOS.}
\begin{tabular}{c c c c c c c c}
 \hline
 Dataset & metric [\%] & BSk20 & DD2 & NL3 & SLy9 & TM1 & \textbf{Avg}  \\ [0.5ex] 

 \hline
 Set 1 &
 \makecell{top-1 \\  top-2 \\ \textit{ds}} & 
 \makecell{99.6 \\ 99.8 \\ -12.0} &
 \makecell{80.8 \\ 87.3 \\ -5.3} &
 \makecell{91.9 \\ 99.8 \\ -4.6} &
 \makecell{47.3 \\ 99.4 \\ -16.2} &
 \makecell{67.6 \\ 97.2 \\ -5.4} &
 \makecell{\bf{77.44} \\ \bf{96.7} \\ \bf{-8.7}} \\

 \hline

 \hline
  Set 2 &
 \makecell{top-1 \\  top-2 \\ \textit{ds}} & 
 \makecell{98.1 \\ 97.2 \\ -4.43} &
 \makecell{54.9 \\ 88.4 \\ -4.58} &
 \makecell{99.8 \\ 100 \\ -1.34} &
 \makecell{24.6 \\ 97.2 \\ -12.84} &
 \makecell{52.5 \\ 98.8 \\ -8.58} &
 \makecell{\bf{65.98} \\ \bf{96.32} \\ \bf{-6.35}} \\

 \hline
 
 \label{table:distance_1_2_other} 
\end{tabular}

\end{center}
\end{table}

The left panel of Figure~\ref{fig:m_1_2_other}
displays the performance of this model on the test set 1, $M_{1,2}/M_{\odot} \in [1.0,1.5]$, which can be compared to that of model of Section \ref{model_1_15}, on the same test set. The corresponding values of the metrics are reported in the top half of Table~\ref{table:distance_1_2_other}. Although both models show good values of the top-1 accuracy, it is seen that for Model C there is a slight decrease in the score (4.14\% averaged across all EOS). Despite this decline being undesired, it can be seen that there is a slight increase in the top-2 accuracy (0.76\% on average) and in the \textit{ds} (2.1\% on average), which compensate the decline in the top-1 accuracy. Therefore, in this region, this model's performance appears to be very similar to that of Model A.

Correspondingly, the right panel of Figure~\ref{fig:m_1_2_other} shows the performance of Model C on the test set 2, $M_{1,2}/M_{\odot} \in [1.5,2.0]$, and the bottom half of Table~\ref{table:distance_1_2_other} reports the values of the associated metrics. While Model C classifies quite well the BSk20 and NL3 EOS (with top-1 accuracy over 90\%), for TM1 and DD2 the numbers go down to 52\% and 55\% respectively and, for SLy9, the model seems to mostly misclassify its samples as BSk20, only achieving a top-1 accuracy of 24\%.
Although these values might seem low, they are a clear improvement over the ones obtained by Model B, with an average rise of 8.9\%. A similar increase is also seen in the top-2 accuracy, going up to an average of 96.32\% (6.14\% more than what is seen for Model B). An additional improvement seen when comparing Table~\ref{table:distance_1_2_other} with Table~\ref{table:distance_15_2} is that the \textit{ds} increases by an average of 3.94\%.

Although the behavior in the lower mass region is very similar to the one displayed by Model A, in the upper mass range $[1.5,2.0]M_{\odot}$ there is a substantial improvement in its performance, over the one exhibited by Model B. This result, together with the high metric values presented in Table~\ref{table:distance_1_2}, shows that training with the full dataset not only leads to a model with good performance across the full parameter range, but it also shows that the resulting model is more capable than its smaller counterparts (when comparing in their respective ranges).

\section{Additional Model\label{sec:additionalmodel}}

The full-range model presented in Sec.~\ref{model_1_2} (Model C) has shown great performance in classifying correctly the EOS  from which the GW signal was generated. Herein, a step forward is given in testing the generalization capacity of this model. This is made by introducing a new EOS, not used in the training stage and using it as input for the model, with the purpose of analysing whether the model classifies the new GW samples as belonging to the nearest EOS in the $\Lambda(M)$ diagram. The EOS added is BSR6 and its $\Lambda(M)$ curve is shown, with a dashed line, in Figure~\ref{fig:mas_plot}. It can be seen that its neighbours $\Lambda(M)$ curves are those of the DD2 and TM1 EOS. Therefore, it is expected that the samples corresponding to BSR6 should be labeled mostly as either one of these two classes.

\begin{figure}[t!]
    \centering
    \includegraphics[width=\linewidth]{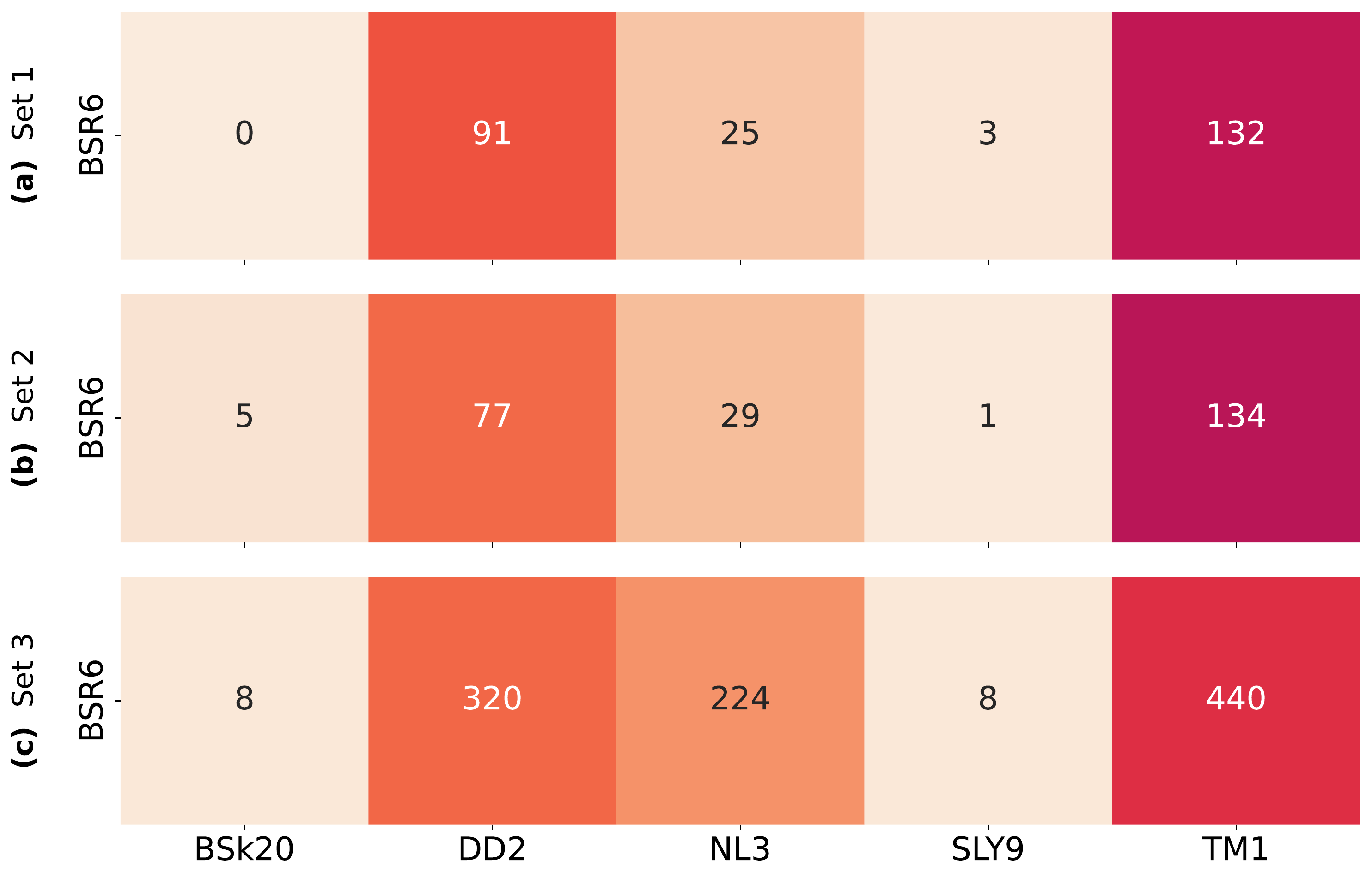} 
    \caption{Confusion matrices of Model C on the BSR6 set 1 (top), $M_{1,2}/M_{\odot} \in [1.0,1.5]$, the BSR6 set 2 (middle), $M_{1,2}/M_{\odot} \in [1.5,2]$, and BSR6 set 3 (bottom), $M_{1,2}/M_{\odot} \in [1.0,2]$.}
    \label{fig:bsr6_confusion}
\end{figure}

The results are shown in Fig.~\ref{fig:bsr6_confusion}, where it can be seen that they are in agreement with the expectation. A clear pattern arises in all three ranges. TM1 is the class predicted in more than half the cases, DD2 follows and, surprisingly, NL3 also has a significant number of predictions. Although, ideally, the predictions would be split only among the two neighboring EOS in the $\Lambda(M)$ diagram (i.e.~DD2 and TM1), the fact that NL3 is commonly predicted is not a sign of failure. The reasoning behind this statement relates to the fact that the  $\Lambda(M)$ curve for NL3 is very similar to that of TM1 (see Fig.~\ref{fig:mas_plot}) which hinders the model's ability to discern between samples from these classes. The confusion between these classes is not new and can be seen in Fig.~\ref{fig:m_1_2_other} where in almost all (95.8\%) instances that the model's prediction fails for samples from TM1, they are misclassified as NL3. 

All in all this test shows that even in the instance of a BNS coalescence event of unknown EOS, the ML model employed in this study is still able to successfully locate it in relation to the EOS employed in its training.

\section{Conclusion} \label{sec:conclusions}

In this work the Audio Spectrogram Transformer (AST) model, a ML model that uses purely an attention-based mechanism, was employed to capture long-range global dependencies for analysing simulated GW data corresponding to BNS coalescences. This is a supervised problem where the goal is to correctly predict the class that each BNS merger event belongs to. For that purpose, a dataset has been generated composed of five classes of BNS mergers that correspond to five distinct equations of state of nuclear matter. The imprint of each EOS on the BNS merger GW inspiral signal is encoded on the tidal deformability dependence on the NS mass, $\Lambda(M)$.

The tidal deformability parameter $\Lambda$ is a steeply decreasing function of the NS mass, covering many orders of magnitude: for a given EOS of the set, $\Lambda$ is ${\cal O}(10^3)$ for $1M_{\odot}$ and ${\cal O}(10)$ for $2M_{\odot}$. However, in the present classification problem, it is the variability of $\Lambda$ among the different EOS for fixed $M$ that most affects the model performance. The difference in $\Lambda$ prediction for the set spans $624< \Delta\Lambda(1M_{\odot})< 5648$ and $74< \Delta\Lambda(2M_{\odot})< 128$, and thus the capacity to distinguish the different EOS strongly decreases with increasing NS mass. Considering this fact, three models were trained on three distinct ranges of NS masses ($M_{\odot}$), namely $[1.0,1.5]$, $[1.5,2.0]$, and $[1.0,2.0]$. Model A showed a very good accuracy, ranging from a top-1 accuracy of 96.5\% for BSk20 to 71.9\% for TM1.
Furthermore, the top-2 accuracy showed that even in the case where misclassifications happened, the true EOS was still the second EOS with highest probability.  
Although Model B presented lower values for the top-1 accuracy (between 84\% and 22\%), the results seen for the top-2 accuracy were still high, with values ranging from 96.2\% to 73.6\%, depending on the EOS. Finally, regarding Model C, when tested on the full dataset it showed that, despite having a significant number of misclassifications, the values of the top-2 accuracy are still very high. More interestingly, when this same model is tested in the range $M_{1,2}/M_{\odot} \in [1.0,1.5]$ it showed similar performance to Model A (see Section \ref{model_1_15}) but, when tested in the range $M_{1,2}/M_{\odot} \in [1.5,2.0]$ it is seen that its results yield a clear improvement over the ones achieved by Model B (see Section \ref{model_15_2}). Therefore, it follows that training in the range $M_{1,2}/M_{\odot} \in [1.0,2.0]$ leads to a robust model that not only preforms well over that range, as expected, but also performs equally well or better in smaller ranges, when compared to models trained specifically for those ranges. 

Despite the fact that this ML approach has been only tested with simulated, noise-free GW data, it already shows great promise. As a result, applications with simulated signals injected into Gaussian noise and real detector noise as well as BNS signals corresponding to  real events will be explored in the future. The preliminary investigations in these directions clearly indicate that a two-model approach may likely be required, i.e.~the present classification model must be supplemented with an already denoised GW signal, and thus a denoising model is required. However, an alternative possibility is exploring different architectures for the AST model that may enable both tasks, denoising and classification, to be carried out simultaneously. This will be the subject of a future work.\\

\section*{Acknowledgements}

F.F.F. is supported by the FCT project PTDC/FIS-PAR/31000/2017 and by the Center for Research and Development in Mathematics and Applications (CIDMA) through FCT, references UIDB/04106/2020 and UIDP/04106/2020. %
M.F. and C.P. acknowledge partial support  by national funds from FCT (Funda\c{c}\~ao para a Ci\^encia e a Tecnologia, I.P, Portugal) under the Projects No. UIDP/-04564/-2020 and No. UIDB/-04564/-2020. %
J.A.F. acknowledges support from the Spanish Agencia Estatal de Investigaci\'on (Grants No. PGC2018-095984-B-I00 and PID2021-125485NB-C21) and from the Generalitat Valenciana (PROMETEO/2019/071).
A.O. acknowledges support from national funds from FCT, under the projects CERN/FIS-PAR/0029/2019 and CERN/FIS-PAR/0037/2021.%
The authors acknowledge the Laboratory for Advanced Computing at the University of Coimbra (http://www.uc.pt/lca) for providing access to the HPC computing resource Navigator,
Minho Advanced Computing Center (MACC) for providing HPC resources that have contributed to the research results reported within this paper, the Portuguese National Network for Advanced Computing for the grant CPCA/A1-428291-2021. %
Finally, the authors gratefully acknowledge the computer resources at Artemisa, funded by the European Union ERDF and Comunitat Valenciana as well as the technical support provided by the Instituto de F\'isica Corpuscular, IFIC (CSIC-UV). %

\bibliographystyle{apsrev}
\bibliography{references}

\end{document}